\documentclass[aip,reprint]{revtex4-1}

\usepackage[pdftex]{graphicx}

\draft 

\begin{document}


\title{Device for the Field Measurements of Frequency-Dependent Soil Properties in the Frequency Range of Lightning Currents}

\author{D. Kuklin}
\email{kuklindima@gmail.com.}
\affiliation{%
Northern Energetics Research Centre, Kola Science Centre of the Russian Academy of Sciences, Apatity, Russia
}%

\date{\today}

\begin{abstract}
Existing approaches for the field measurements of the frequency-dependent soil properties take a significant amount of time, making it difficult to obtain new experimental data and study the electrical soil properties further. However, a relatively uncomplicated measurement device assembled from accessible electronic components can make the measurements almost as easy as those for the regular low-frequency resistivity.
The article presents a detailed description of such a device: the description includes the functional scheme, pseudocodes for central parts of the calculation algorithm, used electronic components, and working principle of the device. The article discusses the reasons behind the chosen parameters of the device (such as isolation between the measurement circuits, generated signal waveform, usage of the electrode arrays) and addresses other possible approaches for the measurements.
The paper also provides several measurement results and a comparison with a calculation result for a particular electrode array configuration. Finally, the work describes essential aspects affecting the accuracy of measurement results.
\end{abstract}

\pacs{}

\maketitle

\section{Introduction}
There is a considerable number of works dedicated to measurements of the complex permittivity of soils, certain minerals, and other similar materials \cite{klein_methods_1997, oh_factors_2007, wang_experimental_2016, datsios_characterization_2019, curtis_durable_2001, huynen_wideband_2001, pauli_versatile_2007, kupfer_tdr_2007, tong_complex_2014, piuzzi_measurement_2016, bore_broadband_2016, bore_large_2017, lewandowski_0053_2017, belyaeva_effect_2016, portela_earth_2006, visacro_frequency_2012}.

Most of the works, however, conduct measurements only with small samples \cite{klein_methods_1997, oh_factors_2007, wang_experimental_2016, datsios_characterization_2019, curtis_durable_2001, huynen_wideband_2001, pauli_versatile_2007, kupfer_tdr_2007, tong_complex_2014, piuzzi_measurement_2016, bore_broadband_2016, bore_large_2017, lewandowski_0053_2017, belyaeva_effect_2016}, which does not represent large volumes of soil (needed for the grounding) as long as most soils are usually inhomogeneous. Besides, a significant portion of works does not cover the frequency range of the lightning currents \cite{curtis_durable_2001, huynen_wideband_2001, pauli_versatile_2007, kupfer_tdr_2007, tong_complex_2014, piuzzi_measurement_2016, bore_broadband_2016, bore_large_2017, lewandowski_0053_2017}.
Summarizing, most of the existing studies solve problems unrelated to the groundings and lightning protection, and they cannot be utilized for the \textit{in situ} measurements of the frequency-dependent soil properties for the frequency range from several kHz (or lower) to several MHz. The current work presents an easy-to-use measurement device that allows conducting this kind of measurements.

There are two approaches (known to the author) for the field measurements of the frequency-dependent soil properties: the first approach performs measurements with large soil samples \cite{portela_earth_2006}, the second approach uses the hemispheric electrode \cite{visacro_frequency_2012}. The method \cite{portela_earth_2006} requires a significant amount of time and work for preparations to measurements as long as it needs collecting relatively large samples from appreciable depth. Additionally, if the soil is inhomogeneous (which is frequently the case), several soil samples should be collected. The method \cite{visacro_frequency_2012} needs much less work but still requires some preparations (related to the hemispheric electrode and remote earth). Besides, all the methods that use at least one common electrode for the current and voltage can potentially lead to inaccurate results due to the electrode polarization and contact resistance between the electrode and surrounding soil (if not used correctly).

A measurement device using electrode arrays can eliminate these drawbacks. The work \cite{kuklin_prototype_2019} presents a prototype of such a measurement device, and the work \cite{kuklin_measurements_2019} provides measurement results of a slightly improved version of the device. These preliminary works pointed out measurement inaccuracies and other problems of the device and suggested several improvements. The present paper describes a measurement device that takes into account the previously proposed improvements. The main differences from the previous preliminary device versions are the significantly improved measurement accuracy and added capability to measure low-frequency resistivity (needed for the calculation of the imaginary part of permittivity).

The article has the following structure: part II discusses the reasons behind the basic parameters of the device, part III presents a detailed description of the device (including the functional scheme and main algorithms), part IV presents several measurement results and a comparison with a calculation result. Additionally, the last part addresses important aspects during measurements.

Instead of the imaginary part of permittivity, this work will mostly use resistivity: these quantities are easily derived from each other when needed. The imaginary part of permittivity, however, is used for the calculations.

\section{Device Parameters and Previous Work}
There is a need to choose several basic parameters for the measurement device: generated waveform, amplitude, type of isolation between the current and voltage measurement circuits, and general design of the device. This part presents the reasoning behind the chosen parameters, describes other tested approaches, and suggests additional possible measurement strategies.

\subsection{Electrode Arrays}
The most critical decision accepted for the device design is that the voltage measurement circuit should be isolated from the generator, which allows using electrode arrays. Usage of the electrode arrays is inherently more accurate than other methods (as it eliminates the influence of the contact resistance), it is convenient (especially if no remote earth is needed) and well understood \cite{loke_recent_2013}, due to longstanding applications in geophysics. However, measurements with the electrode arrays are usually conducted at relatively low frequencies. At the higher frequencies, certain factors should be taken into account to achieve accurate measurement results.

\subsection{Measurement Problems at High Frequencies}
Calculations have shown that for particular array configurations, electromagnetic (EM) coupling effects can lead to significant measurement errors \cite{kuklin_using_2018, kuklin_numerical_2019}. Later, these effects have been observed in measurements, either \cite{kuklin_measurements_2019}. Another important factor is sufficient isolation between the current and voltage measurement circuits. While the EM coupling can be eliminated by choosing appropriate array configurations, isolation between circuits is a more challenging problem at high frequencies. Several different approaches were tested to achieve a proper level of isolation.

The usage of the oscilloscope with insulated channels has shown that parasitic capacitance between the channels does not allow obtaining sufficient isolation.

Then, a measurement device was created, consisting of a generator and two separate blocks that measured current and voltage. Each of these blocks contained an analog-to-digital converter (ADC), and the blocks were isolated with optocouplers (used for the data transmission between the blocks). Due to difficulty in synchronizing two ADCs with accuracy better than one ADC clock cycle, measurement results were inaccurate. Besides, optocouplers still could not provide sufficient isolation.

After that, differential probes were used for the current and voltage measurements. With this approach, measurements of the electrical soil characteristics were successful; however, the probes still have some parasitic capacitance that can influence measurement results.

Finally, an optically insulated voltage probe allowed to achieve perfect isolation between measurement circuits and minimize the error related to the isolation.

\subsection{Main Parameters of the Measurement Device}
Apart from the chosen type of isolation between the measurement circuits, there is a need to make several important decisions concerning other device parameters.

Developing high-quality measurement probes with perfect measurement characteristics over a wide frequency range would make the measurement device unreasonably expensive. However, usage of the calibration makes it possible to achieve accurate measurement results while keeping the measurement probes inexpensive and easy to produce. Dividing the measurement process by two parts (calibration and measurement itself) allows taking into account the imperfections of the amplitude and phase characteristics of the measurement circuits. Therefore this approach was utilized for the measurement device.

Another critical parameter is the generated waveform. Among many possible waveforms, the sine wave seems to be one of the easiest to generate and control: it does not change its form depending on the load (due to low output impedance of the generator) and needs only several microchips to generate a signal with a needed frequency and amplitude. Also, the sine wave allows to perform measurements in the presence of noise: in the case of the periodical signals, the fast Fourier transform (FFT) filters out unwanted noise efficiently, and even a small voltage amplitude is enough for the generated signal; in many cases 10 V amplitude (or less) is sufficient.

The electrical soil properties are calculated based on the measured complex admittance value according to the equation:
\begin{equation}\label{eq:admit}
\hat{Y}(\omega) = \frac{\hat{I}(\omega)}{\hat{V}(\omega)} = k\left(\frac{1}{\rho}+j\omega\epsilon'\right),
\end{equation}
where $k$ is the geometric factor that depends on a particular electrode array \cite{sumner_principles_1976}.
In the case of the sine wave, it is possible to measure the admittance (and properties) in different ways: the measured signals can be either directly digitized by ADC or processed by analog circuits and digitized after that. As an example of the second approach, the phase difference and the amplitude ratio between the current and the voltage can be measured by the chip AD8302. However, this chip needs special techniques to measure phase difference correctly \cite{krok_low-cost_2006}. It is also not very clear how noise in the measured signal would influence measurement results. For the first approach, the noise can lead to errors only in particular cases (see below).

\section{Device Description}
This part provides a detailed description of the device: functional scheme, pseudocodes for the most critical parts of the microcontroller software, used electronic components, smartphone application review.

\begin{figure}
\includegraphics[width=3.3in]{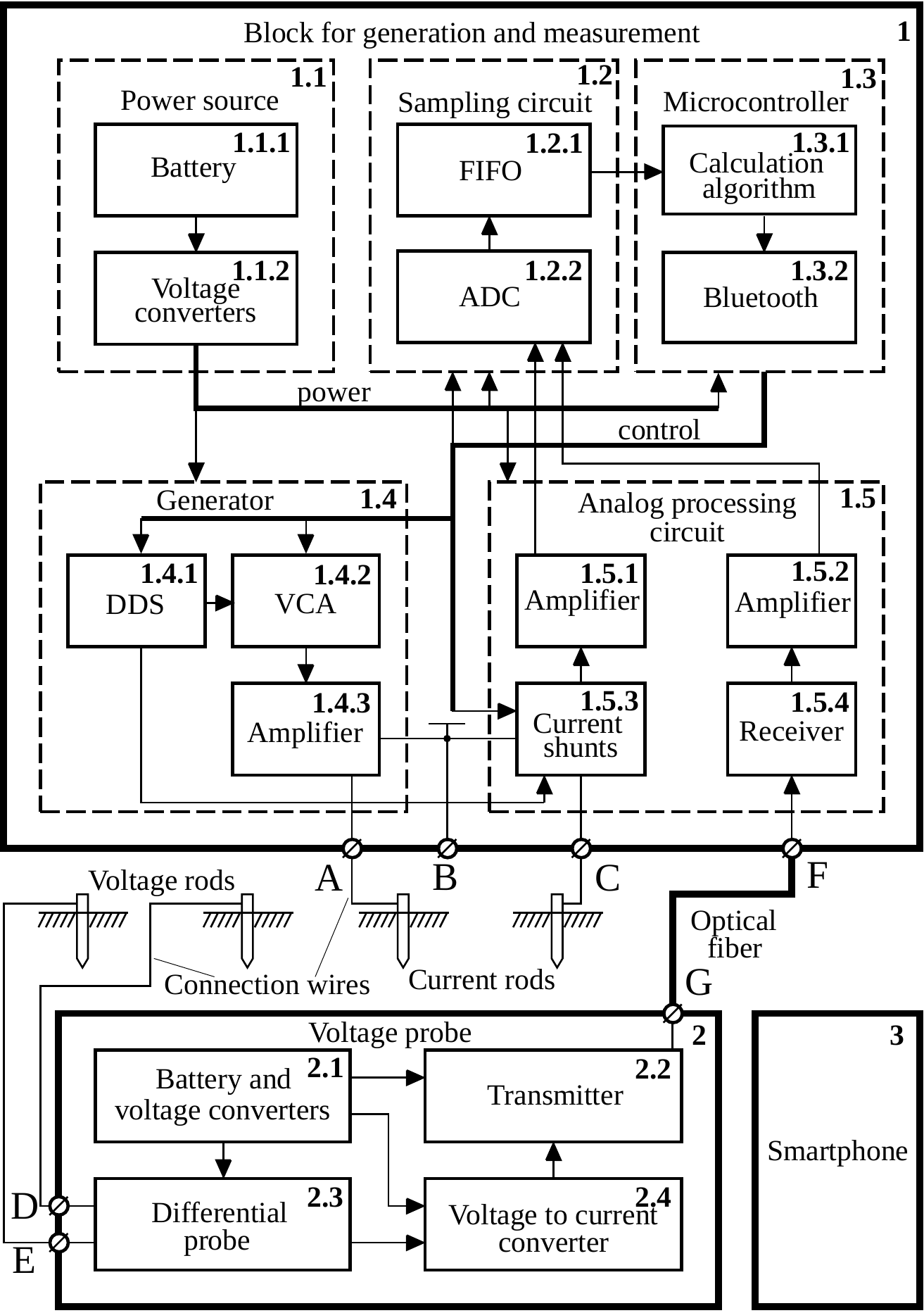}
\caption{Functional scheme of the measurement device.}
\label{fig_func}
\end{figure}

The measurement device consists of three main parts: a block for generation and measurement \textbf{1}, a voltage probe \textbf{2}, and a smartphone \textbf{3} (see Fig.~\ref{fig_func}). The block \textbf{1} is connected between the two current electrodes, and it is responsible for the main functions of the device: digitizing signals, calculation of soil properties, controlling the calibration and measurement. The voltage probe \textbf{2} is connected between the two voltage electrodes. The purpose of the probe is to amplify the measured voltage and convert it to the optical signal. Then the signal is transmitted from the probe \textbf{2} to block \textbf{1} through the optical fiber. The smartphone \textbf{3} is used to control the block \textbf{1} and to perform needed procedures with the measured data.

\begin{figure}
\includegraphics[width=3.4in]{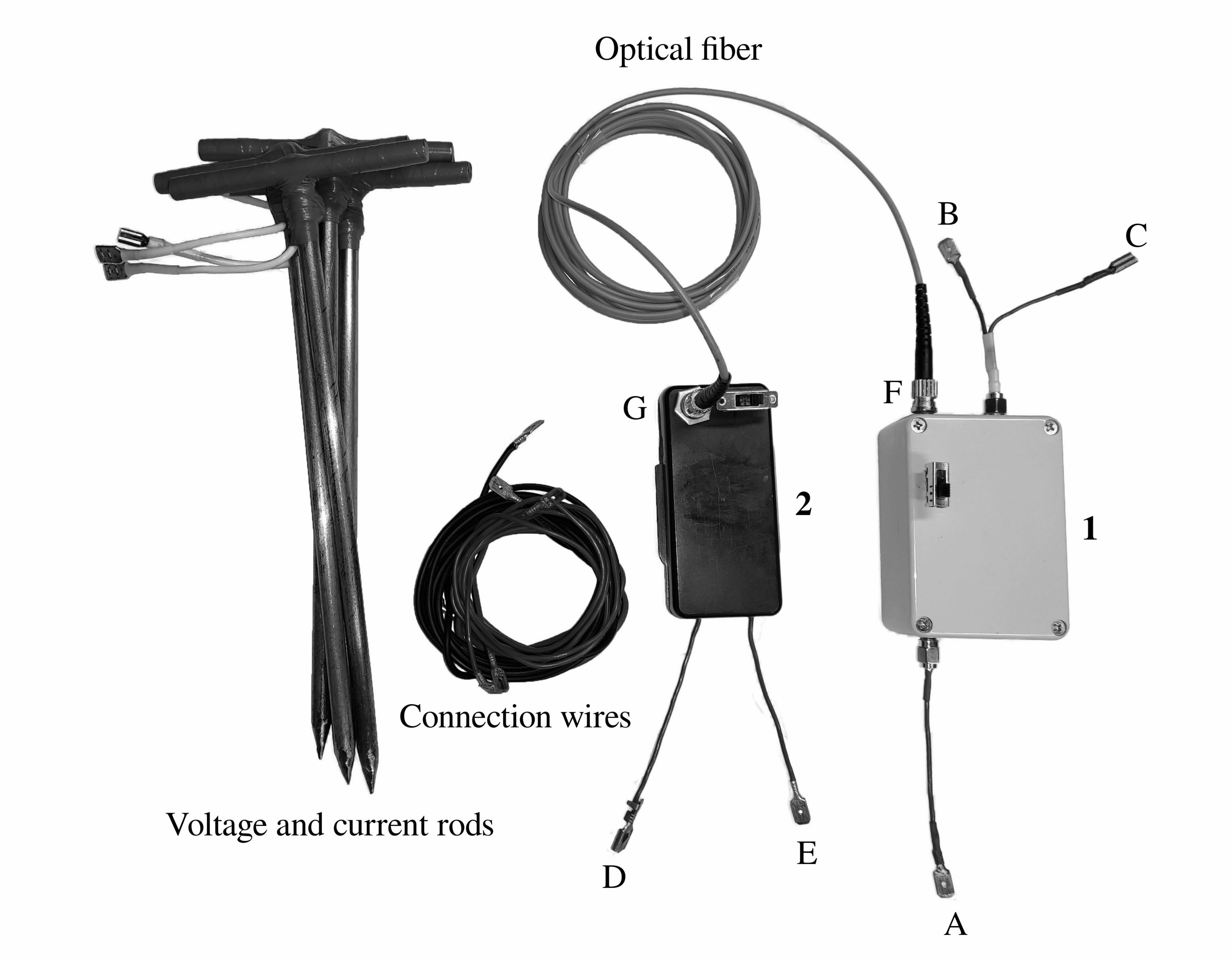}
\caption{View of the measurement device (smartphone is not shown). The captions correspond to those in Fig.~\ref{fig_func}.}
\label{fig_photo}
\end{figure}

Fig.~\ref{fig_photo} shows a picture of the device. The length of the rods in the figure is 25 cm. As the connection wires, regular copper wires with 1 mm$^2$ cross-sectional area were used (unless wires are much thinner, their resistance even at higher frequencies is very low compared to other parameters, such as grounding resistance of the rods or input impedance of the voltage probe). Several measurements with coaxial cables did not lead to improvements in accuracy (although the measurements were not extensive, so it is possible that in other circumstances, the coaxial cable could be preferable).

Housings for the block \textbf{1} and voltage probe \textbf{2} are plastic.

\subsection{Block for Generation and Measurement}
The block for generation and measurement \textbf{1} consists of a power source \textbf{1.1}, sampling circuit \textbf{1.2}, microcontroller \textbf{1.3}, generator \textbf{1.4}, and analog processing circuit \textbf{1.5} (see Fig.~\ref{fig_func}).

The power source \textbf{1.1} consists of a lithium-ion polymer battery \textbf{1.1.1} and several voltage converters \textbf{1.1.2}, which form voltages +3.3 V, +5 V, -5 V, +15 V, and \mbox{-15}~V. The voltage converters use LTC3624 and TPS65131 chips. The battery charger uses the MCP73831 chip. Battery capacity is 2.1 ampere-hours.

The generator \textbf{1.4} consists of direct digital synthesis (DDS) chip AD9834 (\textbf{1.4.1}), voltage-controlled amplifier (VCA) LMH6503 (\textbf{1.4.2}), and operational amplifier LT1210 (\textbf{1.4.3}). VCA allows setting arbitrary amplitude for the generated signal (limited by power source and maximum acceptable supply voltage value for the amplifier \textbf{1.4.3}). The output A is connected to the output of the amplifier LT1210.

The analog processing circuit consists of current shunts \textbf{1.5.3}, receiver HFBR-2416 (\textbf{1.5.4}), and two operational amplifiers THS4503 (\textbf{1.5.1} for the current circuit and \textbf{1.5.2} for the voltage circuit). Current shunts are connected between the ground wire of the device and the output C. Mechanical relays select proper shunt resistance depending on the current through it. The shunts are regular surface-mount device (SMD) resistors (with resistances 10 $\Omega$, 30 $\Omega$, and 100 $\Omega$). The shunts were chosen based on currents' values through them (which depend on the grounding resistance of the current electrodes) and appropriate amplitudes on ADC input after amplification.

\begin{figure}
\includegraphics[width=2.3in]{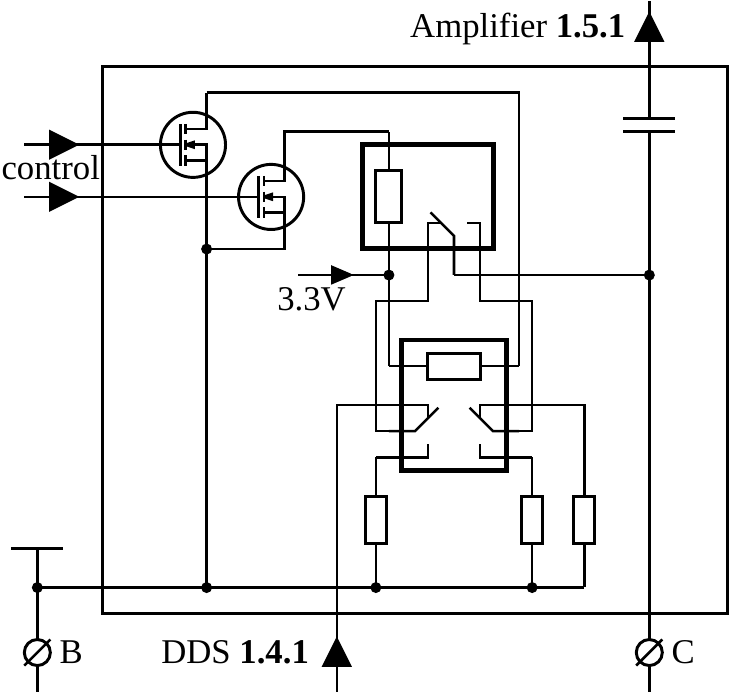}
\caption{Electrical diagram of the block \textbf{1.5.3}.}
\label{fig_shunts}
\end{figure}

Fig.~\ref{fig_shunts} shows a slightly simplified electrical diagram of the block \textbf{1.5.3}. Two relays shown in the diagram allow connecting amplifier \textbf{1.5.1} and output C either to DDS \textbf{1.4.1} (during calibration) or one of the current shunts (during measurements).

The sampling circuit consists of two ADC chips ADS5520 (\textbf{1.2.2}) and two "first-in, first-out" (FIFO) memory chips CY7C4255 (\textbf{1.2.1}). In order to set a clock for ADC very accurately, programmable clock synthesizer CDCE913 is used. Two separate ADC chips allow avoiding crosstalk (which appeared in the previous version of the device \cite{kuklin_measurements_2019} due to the usage of one ADC with multiplexed channels).

The outputs B and C are used for the calibration, during which the sine wave from DDS \textbf{1.4.1} goes to these outputs and current amplifier \textbf{1.5.1}. Thus, to perform the calibration, these outputs should also be connected to the voltage probe (output B directly connects to D, and C directly connects to E). The amplifier \textbf{1.4.3} is turned off during calibration to avoid unwanted noise.

The device uses a microcontroller ESP32 (\textbf{1.3}): apart from the regular capabilities (like serial peripheral interface, inter-integrated circuit, general-purpose input/output, and others), it has integrated Bluetooth and Wi-Fi, which allow controlling the device and transmitting measurement data wirelessly.

The microcontroller software is responsible for the main procedures needed for controlling the measurement process and calculation of the properties. As mentioned above, two main procedures are needed to perform measurements: calibration and the measurement itself. Two functions (executed by the microcontroller) are responsible for these operations.

\begin{figure}
\includegraphics[width=3.4in]{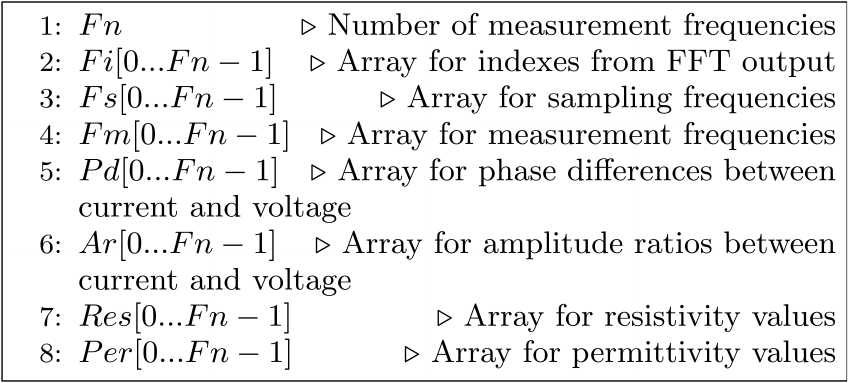}
\caption{Global variables and arrays.}
\label{fig_algdata}
\end{figure}

It is convenient first to describe the main variables and arrays used throughout the code. Fig.~\ref{fig_algdata} presents these variables and arrays. Here, "measurement frequencies" are the frequencies at which resistivity and permittivity values are measured (the DDS chip generates these frequency values), and "sampling frequencies" are the ones used to clock ADC. When FFT is applied to the digitized signals, certain array items (from the FFT output data) should be used. The algorithm ensures that each measurement frequency coincides (approximately) with a particular frequency from FFT output. Array $Fi$ contains values of array items corresponding to the frequencies from FFT output. Arrays $Pd$ and $Ar$ contain calibration data: these are the phases and amplitudes to which voltages (or currents) should be corrected to take into account imperfections of the measurement circuit. Measurement results are stored in the $Res$ and $Per$ arrays.

\begin{figure}
\includegraphics[width=3.4in]{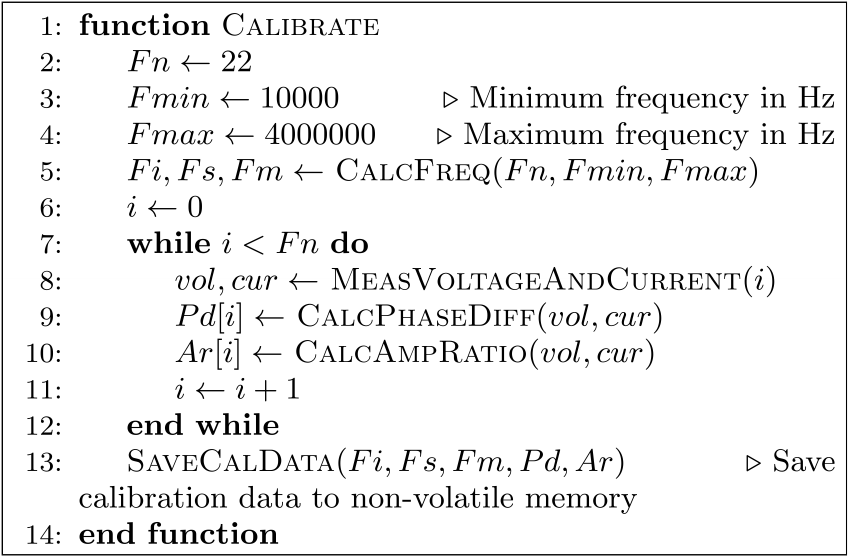}
\caption{Pseudocode for the calibration function.}
\label{fig_algcalib}
\end{figure}

\begin{figure}
\includegraphics[width=3.4in]{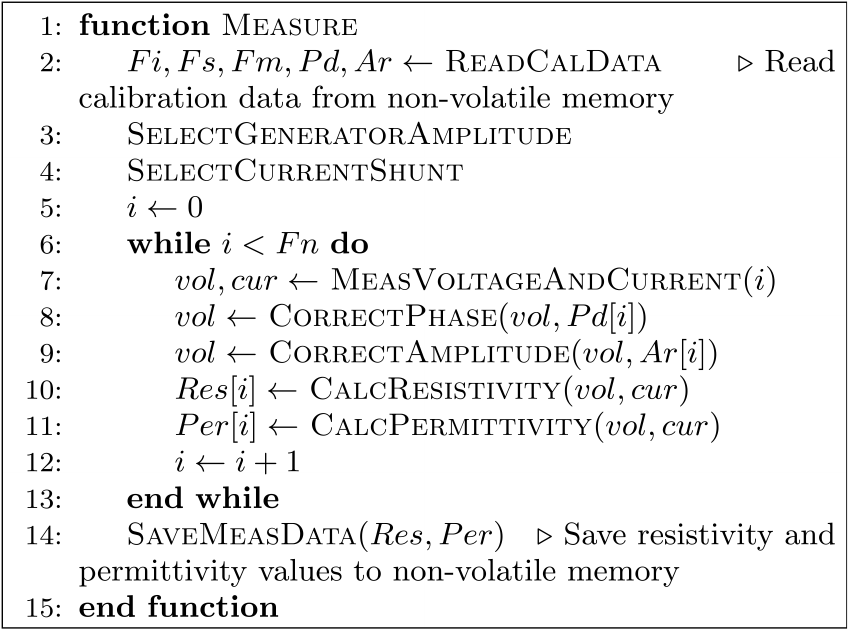}
\caption{Pseudocode for the measurement function.}
\label{fig_algmeas}
\end{figure}

Fig.~\ref{fig_algcalib}, and Fig.~\ref{fig_algmeas} present pseudocodes for the calibration and measurement functions, respectively. Most functions presented here are quite self-descriptive. Also, either because they are simple to implement or very hardware-dependent, only short descriptions are provided for them.

Function \textsc{CalcFreq} is needed for calculations related to the measurement and sampling frequencies: first, the frequencies should be chosen so that each measurement frequency coincides with one of the FFT output frequency (as mentioned above), and second, measurement frequencies should be placed linearly on the logarithmic scale.

During calibration, the same signal goes to the voltage probe and the current measurement circuit. As long as the signal goes through two circuits with different amplitude-frequency and phase-frequency characteristics, two signals at the circuit outputs have frequency-dependent phase difference and amplitude ratio. Functions \textsc{CalcPhaseDiff} and \textsc{CalcAmpRatio} calculate phase difference and amplitude ratio for two phasors (or, in other words, for two signals measured by ADC).

Functions \textsc{SaveCalData} and \textsc{SaveMeasData} save data to flash memory, and function \textsc{ReadCalData} reads data from memory. Their implementation depends on the chosen microcontroller, and usually, it is a typical procedure.

Functions \textsc{SelectGeneratorAmplitude} and \textsc{SelectCurrentShunt} automatically choose the current shunt and generator amplitude, which was proposed in the previous work \cite{kuklin_measurements_2019}. These functions choose the needed parameters based on measured current and voltage amplitudes before each measurement.

Functions \textsc{CorrectPhase} and \textsc{CorrectAmplitude} correct voltage phasor to compensate phase and amplitude errors (determined during calibration).

Functions \textsc{CalcResistivity} and \textsc{CalcPermittivity} calculate resistivity and permittivity from voltage and current according to equation (\ref{eq:admit}). The properties are calculated for $k$~=~1 and transmitted to a smartphone, where the actual value of $k$ is calculated based on electrode coordinates set in the smartphone application.

\begin{figure}
\includegraphics[width=3.4in]{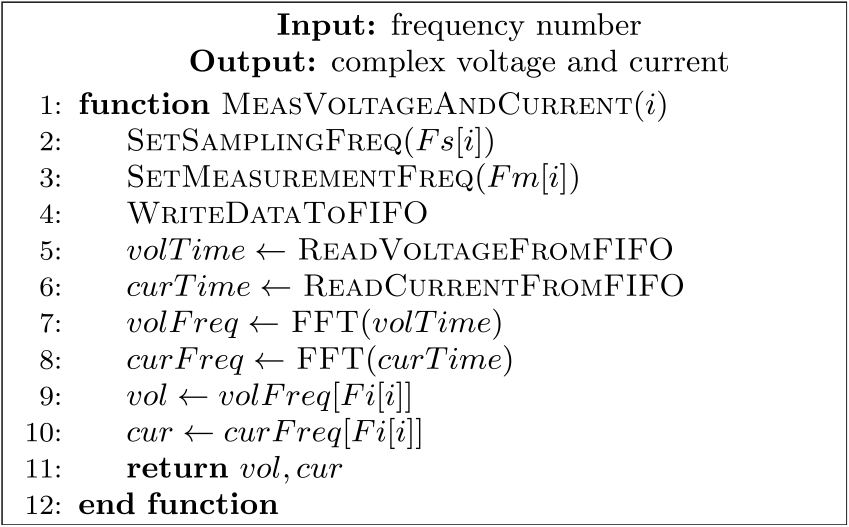}
\caption{Function for measurement of the complex current and voltage.}
\label{fig_algcurvol}
\end{figure}

The calibration and measurement functions use the common function \textsc{MeasVoltageAndCurrent}, shown in Fig.~\ref{fig_algcurvol}. Functions \textsc{SetSamplingFreq}, \textsc{SetMeasurementFreq}, \textsc{WriteDataToFIFO}, \textsc{ReadVoltageFromFIFO}, and \textsc{ReadCurrentFromFIFO} are hardware-dependent. \textsc{SetSamplingFreq} is used to set a certain frequency for the CDCE913 output (ADC clock input). \textsc{SetMeasurementFreq} sets the frequency for DDS chip AD9834. \textsc{WriteDataToFIFO} performs all the needed actions for turning on ADC and writing data to FIFO. Functions \textsc{ReadVoltageFromFIFO} and \textsc{ReadCurrentFromFIFO} are needed to transfer data from FIFO to internal random-access memory (RAM) of the microcontroller so that \textsc{FFT} function could apply fast Fourier transform to the time-domain data. After that, $Fi$ array is used for accessing certain indexes of $volFreq$ and $curFreq$ arrays so that needed complex voltage and current values could be finally obtained.

Apart from the measurements for the frequency range specified by the $Fmin$ and $Fmax$, the device also measures resistivity at a relatively low frequency (500~Hz in this case), the value of which is limited by minimum ADC sampling frequency and FIFO size. This value is needed for the calculation of the imaginary part of permittivity.

Currently, almost no special measures are taken to increase accuracy (algorithmically). However, there are some possible ways to do that. For example, some values can be measured several times (to calculate the average value). Also, certain types of noise (see below) could be eliminated by an algorithm.

Calibration takes around 30 seconds, and measurement needs the same amount of time (for 32 frequencies). I.e., it takes around 1 second per frequency (independently on measured values). Thus, preparation for the measurements (such as the accurate location of the electrodes, for example) usually takes more time than the measurements themselves.

\subsection{Voltage Probe}
The voltage probe \textbf{2} consists of a 1 ampere-hour lithium-ion polymer battery and voltage converters \textbf{2.1}, differential probe \textbf{2.3}, a voltage to current converter \textbf{2.4}, and transmitter \textbf{2.2}. The voltage converters form voltages +5 V (RT9266) and -5 V (MAX660). The differential probe \textbf{2.3} is based on two junction gate field-effect transistors (JFET) SST310 and operational amplifier LT1807. Voltage to current converter \textbf{2.4} consists of an operational amplifier LT1807 and a bipolar junction transistor (BJT) MMBT3904. For the transmitter \textbf{2.2}, HFBR-1414 is used.

As the optical fiber, 62.5/125 (core/cladding diameter) fiber patch cord is used. The straight tip (ST) connectors provide a convenient way to replace the cable if required.

The input resistance of the voltage probe is higher than 20 M$\Omega$; input capacitance is lower than 0.55 pF.

\subsection{Smartphone Application Software}
The device needs means for controlling it, displaying measurement results, saving the results, and others. It is possible to embed these capabilities in the device itself. However, a regular smartphone has all the needed functions using which it is possible to simplify the measurement device significantly. Moreover, creating a smartphone application is usually a more straightforward task than creating a graphical user interface for a touch screen display controlled by a microcontroller. Another advantage is that remote control allows avoiding the measurement error caused by accidental touching of some parts of the device (due to parasitic impedance of the human body). Thus, it was chosen to create a smartphone application for controlling the device.

\begin{figure}
\includegraphics[width=2.3in]{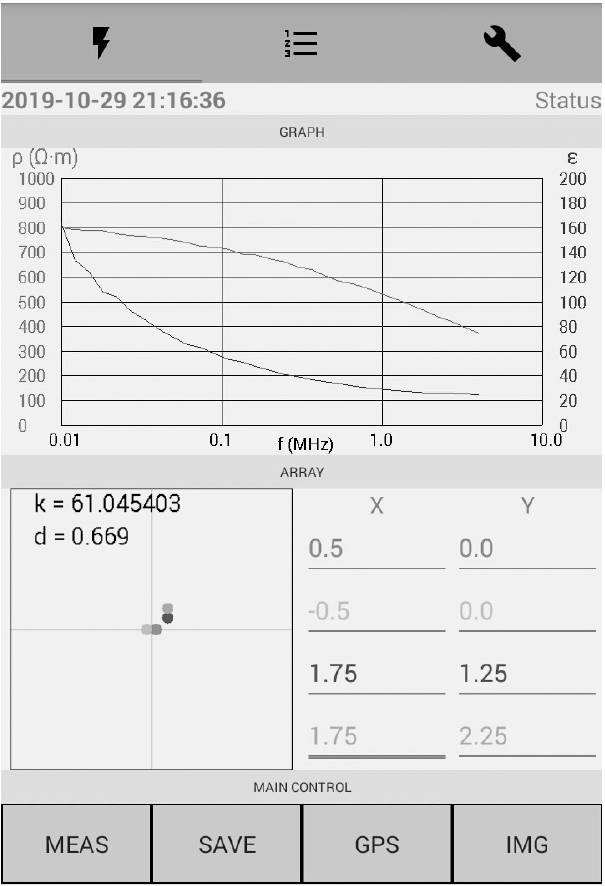}
\caption{View of the smartphone application.}
\label{fig_app}
\end{figure}

Fig.~\ref{fig_app} shows the view of the app. The app connects to the measurement device via Bluetooth and allows triggering calibration and measurement processes. It also receives measurement results from the device and saves them (and can display the results as a graph). Also, because most smartphones have a camera and a global positioning system (GPS) receiver, the app allows taking pictures of the measurement site and saving geographic coordinates of the place. Additionally, it is possible to view voltages digitized by ADC to ensure that there is no noise or other factors that can lead to erroneous results. Apart from automatic generator amplitude and current shunt selection (by device), manual selection of these parameters is possible from the application. In order to calculate soil properties correctly, the app allows setting coordinates of the current and voltage electrodes: the app uses the coordinates to calculate the geometric factor $k$ and the depth of investigation for an array. The geometric factor is calculated as:
\begin{equation}\label{eq:geomf}
k = \frac{2\pi}{\frac{1}{r_1}-\frac{1}{r_2}-\frac{1}{R_1}+\frac{1}{R_2}},
\end{equation}
where $r_1$ and $r_2$ are the distances from the first potential electrode to the first and the second current electrodes, $R_1$ and $R_2$ are the distances from the second potential electrode to the two current electrodes. The electrodes' position relative to axes (in the app) is not important, as long as the mutual location of the electrodes corresponds to those in measurements.

Currently, the device itself calculates the needed properties. Another option is to send measured voltages and currents to the smartphone and calculate the properties inside the smartphone. This approach can reduce the time for measurement if a microcontroller is not very efficient. For faster data transmission speed, Wi-Fi can be used. However, this can reduce battery life due to higher power consumption.

\section{Measurements}
This part presents several measurement results and a comparison with a calculation result. The part also addresses several factors affecting the accuracy of measurement results.

\begin{figure}
\includegraphics[width=2.5in]{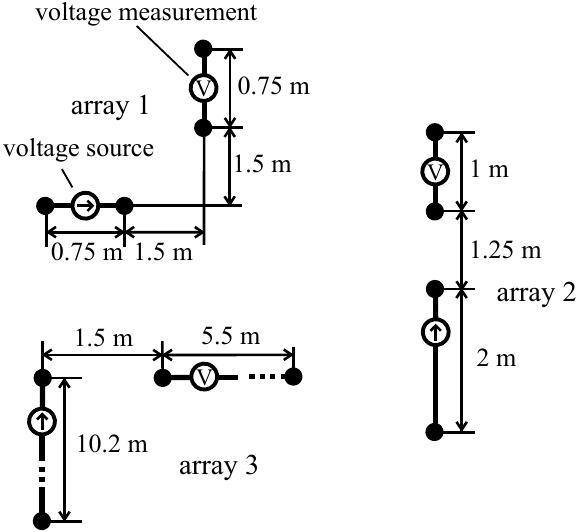}
\caption{Electrode arrays used in measurements.}
\label{fig_arrays}
\end{figure}

Fig.~\ref{fig_arrays} illustrates a top view of electrode arrays used for the measurements.

\begin{figure}
\includegraphics[width=3.4in]{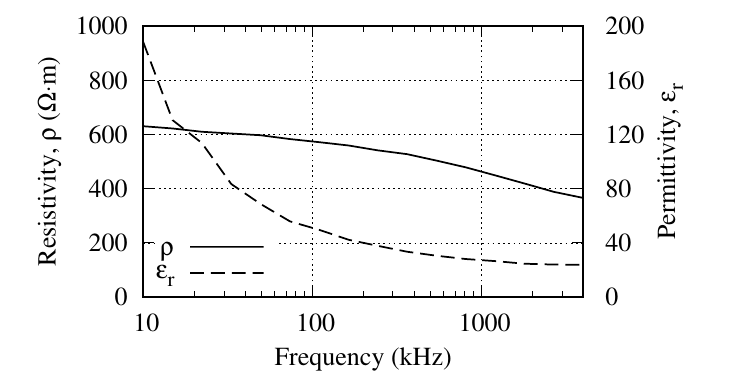}
\caption{Measurement result for the array 1.}
\label{fig_array1}
\end{figure}

\begin{figure}
\includegraphics[width=3.0in]{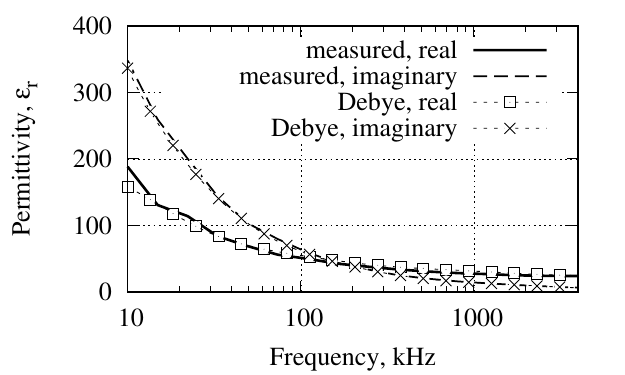}
\caption{Real and imaginary parts of measured permittivity and their approximation.}
\label{fig_array1eps}
\end{figure}

Fig.~\ref{fig_array1} shows the measurement result for the array 1. The value of low-frequency resistivity $\rho_{0}$, in this case, is 718.76~$\Omega\cdot$m. Fig.~\ref{fig_array1eps} shows the same result expressed as complex permittivity. This figure also shows real and imaginary permittivity approximated with the Debye relaxation model (needed for the calculations below).

Calculations and measurements for the array 2 allow verifying if there is an agreement between the calculation model and measurement results.

\begin{figure}
\includegraphics[width=2.5in]{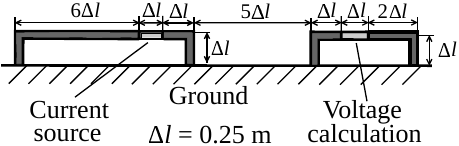}
\caption{Calculation model for the array 2 ($\Delta l$ is the calculation grid cell size).}
\label{fig_arraysside}
\end{figure}

Calculations are made with the finite difference time domain (FDTD) method \cite{yee_numerical_1966, taflove_numerical_1975, taflove_computational_2005}, which uses the auxiliary differential equation method \cite{taflove_computational_2005,okoniewski_simple_1997} to model Debye relaxation (and, subsequently, frequency-dependent soil properties). The Debye function expansion:
\begin{equation}\label{eq:debye}
\hat{\epsilon}_r(\omega) = \epsilon_\infty + \sum_{p=1}^{n} \frac{\Delta\epsilon_p}{1+j\omega\tau_p}.
\end{equation}
where $n$ is the number of Debye terms (poles). Table~\ref{table_param} shows the parameters $\Delta\epsilon$ and $\tau$ of the expansion calculated with the hybrid particle swarm-least squares optimization approach \cite{kelley_debye_2007}. $\epsilon_{\infty}$ equals to 22.979.
Calculation parameters, such as current function \cite{heidler_class_2002}, absorbing boundary conditions \cite{taflove_computational_2005}, the method for modeling thin wires \cite{railton_treatment_2005}, are the same as previously \cite{kuklin_numerical_2019}. Fig.~\ref{fig_arraysside} illustrates the calculation model.

\begin{table}
\caption{\label{table_param}Parameters for the Four-Term Debye Function Expansion}
\begin{ruledtabular}
\begin{tabular}{ccc}
term & $\Delta\epsilon$ & $\tau, s$\\
\hline
1 & $11.100$ & $1.102\cdot10^{-7}$\\
2 & $24.950$ & $1.069\cdot10^{-6}$\\
3 & $137.225$ & $9.984\cdot10^{-6}$\\
4 & $1.102\cdot10^{6}$ & $6.420 \cdot10^{-2}$\\
\end{tabular}
\end{ruledtabular}
\end{table}

The soil properties, measured with the array 1, can be used in calculations with array 2 (measurements with arrays 1 and 2 were made in the vicinity to each other to ensure that the soil properties are approximately the same). After that, a comparison can be made with measurement results for the array 2.

\begin{figure}
\includegraphics[width=3.4in]{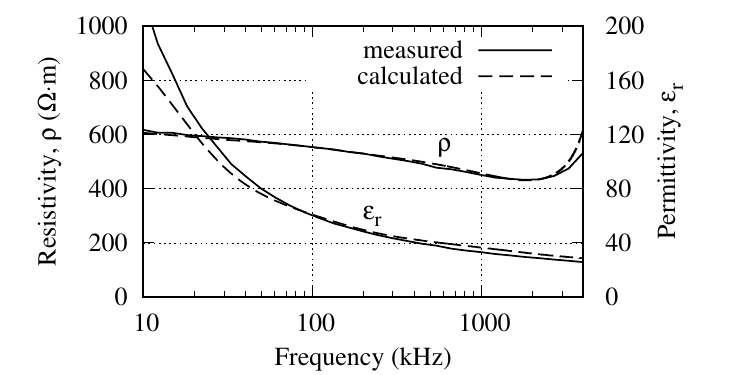}
\caption{Measurement and calculation results for the array 2.}
\label{fig_array2}
\end{figure}

Fig.~\ref{fig_array2} shows measurement and calculation results for the array 2. Taking into account that the soil is inhomogeneous and different arrays have different sensitivity patterns, this result demonstrates that there are same EM coupling effects in measurements and calculations. The results also provide additional validation to the calculation model.

\begin{figure}
\includegraphics[width=3.4in]{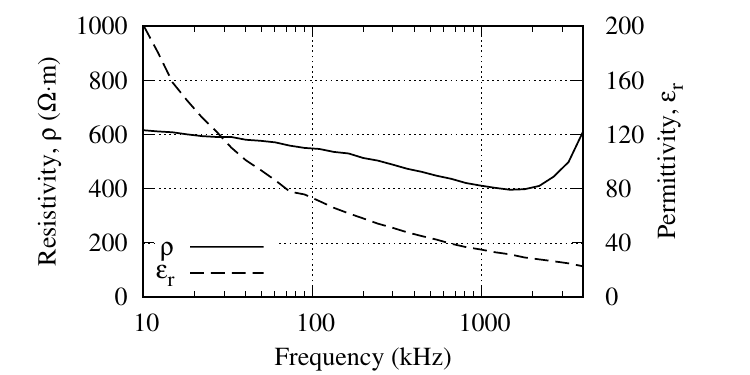}
\caption{Measurement result for the array 3.}
\label{fig_array3}
\end{figure}

Previous works have shown that the measurement errors can exist even for arrays with perpendicularly located measurement wires \cite{kuklin_measurements_2019, kuklin_numerical_2019}. However, the current device version is more accurate (due to the amplitude correction during measurements, absence of the crosstalk, and better temperature stability of the voltage probe). Therefore similar measurements were repeated here (see array 3 in Fig.~\ref{fig_arrays}). Measurement results for the array 3 (shown in Fig.~\ref{fig_array3}) confirm the previous results.

It should be mentioned that several preliminary measurements have shown that using the connection wires during calibration helps to reduce the measurement error. Thus, it is possible that long connection wires could be a significant contributor to the resistivity measurement error for arrays with perpendicular dipoles. As long as the measurements were preliminary, this should be examined by additional measurements with different arrays and, if possible, verified by calculations.

\subsection{Important Aspects During Calibration}
As mentioned previously, there are two parameters measured during calibration: phase difference and amplitude ratio.
Accurate phase calibration is crucial: even a small error in angle due to noise or other factors can lead to significant errors during measurements. Besides, the phase difference depends on frequency quite noticeably. Amplitude is usually more uniform throughout the frequency range than the phase (with correctly designed amplifying circuits). However, uncorrected amplitude still can introduce some error: comparing results with no amplitude correction \cite{kuklin_measurements_2019} and results with the correction (above), one can see the noticeable difference.

The voltage probe can be very sensitive to the external EM field: when the device was tested, unwanted signal several times higher than the calibration signal (but not connected to the calibration outputs electrically) was able to cause significant phase and amplitude errors (and measurement errors later). Because of the sensitivity, usage of the differential probe for current measurements can cause errors (due to its proximity to the generator).

Another factor that can influence measurement accuracy is the temperature of the measurement device (during calibration and measurement). The device temperature can change either due to ambient temperature or because of the heating of electronic components (during their regular operation). And if the temperature during a measurement is different from one during calibration, this can lead to measurement error. Thus, either the temperature dependence should be minimized, or some temperature compensation should be used (to take into account the influence of temperature on measurement results). Also, to increase measurement accuracy, calibration could be done each time immediately before measurement.

One should also avoid significant bending of the optical fiber during calibration, as long as this influences the amplitude of the signal.

\subsection{Important Aspects During Measurements}
The most critical parameter for the measurements, as mentioned previously, is the location of connection wires: to reduce EM coupling, the wires should be located perpendicularly.

Another parameter is the length of measurement wires: they should be relatively short (preferably, not longer than several meters). Short distances, however, increase geometric factor $k$ and, therefore, can lead to errors due to small (compared to noise) amplitude of the measured voltage. Thus, distances should not be too short either (unless a small depth of investigation or a specific sensitivity pattern is needed). Both current and voltage rod distance influence EM coupling error; the noise level is mostly influenced by the voltage rod distance \cite{kuklin_measurements_2019} and location of the voltage rods in general.

Usually, FFT effectively filters out high-frequency (and relatively low-amplitude) noise. In some cases, however, low-frequency noise with relatively high amplitude was observed (even for a small distance between voltage electrodes). This noise was able to shift the useful signal beyond allowable ADC input voltage range. This kind of error could be eliminated by detecting erroneous signals algorithmically (and repeating measurements) or by interchanging current and voltage rods during measurements.

There are also some other factors that can slightly influence measurement results at high frequencies, such as generator location along the line between current rods and its elevation above the ground.

\begin{table}
\caption{\label{table_paramdev}Parameters of the Measurement Device}
\begin{ruledtabular}
\begin{tabular}{ccc}
parameter & value\\
\hline
Resistance measurement range & 100 $\Omega$ ... 10 k$\Omega$\\
Capacitance measurement range & 47 pF ... 2.7 nF\\
Resistance accuracy (below 2 MHz) & $\approx$4\%\\
Resistance accuracy (above 2 MHz) & $\approx$10\%\\
Capacitance accuracy (above 30 kHz) & $\approx$6\%\\
Capacitance accuracy (below 30 kHz) & $\approx$10\%\\
Frequency range & 10 kHz ... 4 MHz\\
Frequency accuracy & 0.28 Hz\\
Signal amplitude (current circuit) & 4 mV ... 110 mV\\
Signal amplitude (voltage circuit) & 4 mV ... 140 mV\\
Sensitivity (current circuit) & $\pm$ 12.6 mV\\
Sensitivity (voltage circuit) & $\pm$ 30.3 mV\\
\end{tabular}
\end{ruledtabular}
\end{table}

Below, several parameters of the measurement device are given (see Table~\ref{table_paramdev}). There are challenges in determining the measurement range and accuracy of the measurement device as it is difficult to find reference measurement results (analogous to those in the field) for comparison. A possible approach (used in this work) is to measure resistivities and capacitances and compare results to their known values. However, there are significant differences between the field measurements and measurements with lumped elements. Thus, parameters in Table~\ref{table_paramdev} are only estimations (actual measurement range and accuracy for the field measurements can be different). Possibly, there is a more accurate way to determine these parameters. One of the indicators of accuracy for the field measurements is the good agreement between the measured complex permittivity and the Debye relaxation model (for which real and imaginary parts of the permittivity related by the Kramers-Kronig relations).

\begin{figure}
\includegraphics[width=1.4in]{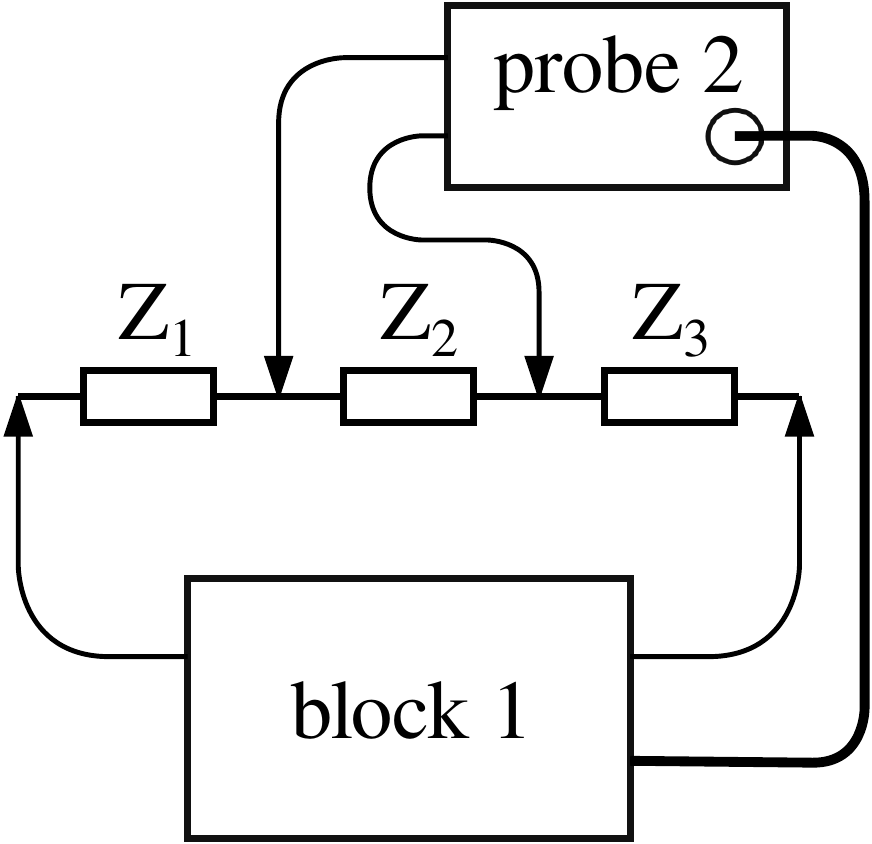}
\caption{Connection to voltage divider.}
\label{fig_divider}
\end{figure}

In order to test the measurement range and accuracy, a voltage divider was used (see Fig.~\ref{fig_divider}). As the impedances $Z_1$, $Z_2$, and $Z_3$, either resistors or capacitors were used. When a capacitor and resistor are connected in parallel, in some cases there are errors significantly exceeding those during the field measurements. Thus, results in the table were measured for resistors and capacitors separately. Also, to avoid high amplitudes at ADC input, voltage probe amplification was reduced, and VCA output was used as generator output. In addition, measured values seemingly depend on the amplitude of signals (which affects measurement accuracy); therefore, high amplitudes were avoided. This amplitude dependence can be taken into account in the next version of the device.

The frequency range in the table corresponds to that used in the present work. It is limited mostly by ADC clock and FIFO memory (also partially by the voltage probe and generator), and technically it can be wider: starting from several hundreds of Hz till around 10 MHz. However, resistivity measurement accuracy at high frequencies and permittivity accuracy at low frequencies would be low in that case. But low-frequency resistivity measurement is important, as it was mentioned above.

According to the AD9834 datasheet, this DDS chip allows achieving a resolution of 0.28 Hz (with 75 MHz clock), which determines the frequency accuracy of the generator.

Amplifier \textbf{1.5.1} amplifies the signal by 10; amplifier \textbf{1.5.2} and the voltage probe amplify the signal approximately by 8 (depending on frequency). As long as the ADC input range is 2.3 V, this determines maximum amplitudes for the voltage probe and voltages on the current shunts. Minimum amplitudes were determined by gradual reduction of calibration signal until visible errors started to appear. The minimum amplitudes can be lower or higher depending on the acceptable level of measurement errors (thus, this is an approximate value).

Peak-to-peak sensitivity was calculated as $\pm N$ times 2.3 V divided by 4096, where $\pm N$ corresponds to the noise measured in counts, 2.3 V is the ADC input range, 4096 is $2^{12}$ (the ADC has 12 bits resolution). Note that the noise was measured at ADC input, where the signal is amplified by the voltage probe and the analog processing circuit \textbf{1.5}. In addition, FFT allows detecting weak signals even in the presence of noise (amplitudes lower than noise probably can be detected).

\section{Conclusion}
The measurement device presented in the article significantly decreases measurement time (down to 10--20 minutes or so) and improves the convenience of use due to its compact size, making field measurements of frequency-dependent soil properties very easy. Furthermore, the device does not contain very specific electronic components (and does not need special programming skills); therefore, it is relatively easy to replicate the device. Thanks to this, more field experimental results can be obtained, so that the frequency-dependence of soil properties can be investigated more thoroughly. The device could even be used for engineering purposes if the electrical exploration of soil is needed (similarly to the regular low-frequency resistivity measurements). Possibly, the modified version of the device could also be used for measurements with soil samples.

In the cases when larger investigation depths are needed, generator amplitude should be increased.

As long as soils are usually inhomogeneous, more complicated soil models could be needed \cite{datsios_methods_2020, li_inversion_2020}. However, it probably would be appropriate first to determine what accuracy (for inhomogeneities) is needed in practice. It is possible that, in particular cases, it is not very important to know all the inhomogeneities of soil. I.e., it could be enough to know some "effective" soil properties for a volume of soil where grounding is (or will be) located. And possibly those "effective" properties could be measured by the device.

\bibliography{references}

\section{Data Availability Statement}
The data that support the findings of this study are available from the corresponding author upon reasonable request.

\section{AIP Publishing credit line for Accepted Manuscript}
This article may be downloaded for personal use only. Any other use requires prior permission of the author and AIP Publishing. This article appeared in Review of Scientific Instruments 91, 114701 (2020) and may be found at https://aip.scitation.org/doi/10.1063/5.0012126

\end{document}